\documentclass[12pt]{article}
\usepackage{geometry}
\usepackage{fancyvrb}
\usepackage{lipsum}
\usepackage[width=.8\textwidth]{caption}

\usepackage{graphicx}
\usepackage{amsmath,amsthm,amssymb}
\usepackage{enumerate}
\usepackage{comment}
\usepackage{grffile}
\usepackage{amsfonts}
\usepackage{amsthm}
\usepackage{cases}
\usepackage{color}
\usepackage{setspace}
\usepackage{float}
\usepackage{booktabs}
\usepackage{array}
\usepackage{subfigure}
\usepackage{bm}
\usepackage{booktabs}
\usepackage{array}
\usepackage{multirow}
\usepackage{comment}
\usepackage{url}
\usepackage[sort&compress]{natbib}
\usepackage{hyperref}
\hypersetup{
    colorlinks=black,
    linkcolor=blue,
    filecolor=magenta,
    urlcolor=cyan,
    citecolor =black,
}

\begin{document}
\begin{center}\textbf{\Large{A Clustering Approach to Integrative Analysis of Multiomic Cancer Data}}\\
\large{Dongyan Yan} \\
\large{Formerly at University of Missouri, currently employed at Discovery \& Development Statistics, Eli Lilly and Company} \\
\large{and} \\
\large{Subharup Guha} \\
\large{Department of Biostatistics, University of Florida}
\end{center}

\textbf{Abstract:} Rapid technological advances have allowed for molecular profiling across multiple omics domains from a single sample for clinical decision making in many diseases, especially cancer. As tumor development and progression are dynamic biological processes involving composite genomic aberrations, key challenges are to effectively assimilate information from these domains to identify genomic signatures and biological entities that are druggable, develop accurate risk prediction profiles for future patients, and identify novel patient subgroups for tailored therapy and monitoring. 
\\
 We propose integrative probabilistic frameworks for high-dimensional multiple-domain cancer data that coherently incorporate dependence within and between domains to accurately detect tumor subtypes, thus providing a catalogue of genomic aberrations associated with cancer taxonomy. We propose an innovative, flexible and scalable Bayesian nonparametric framework for simultaneous clustering of both tumor samples and genomic probes. We describe an  efficient variable selection procedure to identify relevant genomic aberrations that can potentially reveal underlying drivers of a disease. Although the work is motivated by several investigations related to lung cancer, the proposed methods  are broadly applicable in a variety of contexts involving high-dimensional data. The success of the  methodology is demonstrated using artificial data and lung cancer omics profiles publicly available from The Cancer Genome Atlas. \\

\section{Introduction}
An overarching goal of cancer omics studies is identifying patient  profiles based on various genetic and epigenetic aberrations in tumors \citep{simon2013implementing, navin2014cancer, wang2018identification}. 
Rapid technological advances in recent years have enabled molecular profiling of  patients across multiple omics domains, such as SNP, mRNA,  gene and protein expression.   In the past,  studies have typically analyzed the somatic measurements one domain at a time with some degree of success. For example,  high-resolution RNA sequencing and microRNA profiling have led to the discovery of novel cancer molecular subtypes in bladder cancer \citep{choi2017genetic} and  breast cancer  \citep{haakensen2016subtype}. 
However,
as tumor development and progression involve complex genomic aberrations occurring across multiple domains, it is important,  although statistically challenging, to effectively assimilate information across the domains to accurately identify biological entities,  genomic signatures,  and  risk prediction profiles for the patients. This is especially true of  ``small $n$, large $p$'' problems where the number of biomarkers overwhelms the number of patients.

\subsection{Motivating application}
The Cancer Genome Atlas (TCGA) is a worldwide project that focuses on cataloging genetic mutations responsible for various cancers   (\path{https://www.cancer.gov/about-nci/organization/ccg/research/structural-genomics/tcga}). 
 Currently, TCGA characterizes more than 30 cancer types,   including 10 rare cancers. 
The advent of TCGA has further incentivized the integration of data from multiple domains to identify meaningful and druggable biomarkers associated with tumor development and progression. 

Lung cancer is the second most common cancer in both men and women in the US, to date accounting for more than 235,760 new cases and 131,880 deaths in 2021 (\path{https://www.cancer.org/cancer/lung-cancer/about/key-statistics.html}). It is the number one cause of cancer death among  men and women, constituting almost 25\% of all cancer deaths. Lung adenocarcinoma (LUAD) and lung squamous cell carcinoma (LUSC) are the two major subtypes of non-small cell lung cancer (NSCLC), which accounts for more than 80\%  of all lung cancers  \citep{singh2020comprehensive}. A comprehensive genomic profiling on LUSC and LUAD tumor samples has been studied in \cite{cancer2014comprehensive, cancer2012comprehensive}. Although we focus on the integrative analysis of gene expression and DNA methylation data in Section \ref{subsection: TCGA}, but the proposed methodology is applicable to an arbitrary number of platforms. 

\subsection{Existing methods}
Existing integrative methods can be  categorized into {four} groups based on the main study goals. The first group involves one-at-a-time analysis of heterogeneous data from different platforms to validate biological assumptions. These studies obtain data from one platform, analyze it along with outcomes of interest, and then perform a similar analysis using a different platform for the purpose of validating the results of the previous analysis. For example, \cite{cenik2015integrative} used ribosome occupancy maps coupled with previously measured RNA expression and protein levels from the same group of individuals to improve the identification of disease risk factors. \cite{lian2019integrative}  obtained one stemness index from gene expression, another stemness index from DNA methylation for verification, and observed an inverse trend between them for metastatic status.

The second group of studies focuses on the discovery and characterization of biological pathways  or  regulatory mechanisms associated with the disease. For example, \cite{wu2018integrative} studied the associations among DNA methylation and phenotype for follow-up functional studies to reveal complex disease traits. \cite{yoo2019integrative} and \cite{fan2019integrative} performed  integrative analyses of two or more omics platforms to detect  possible signaling pathways related to the progression of different cancers. 

The third group of studies aim to capture biological heterogeneity from various platforms to  detect sparse subsets of informative variables that predict clinical outcomes. For example, {\cite{wang2013ibag} proposed a hierarchical model to capture the natural mechanistic relationships among multi-platform genomic data, thereby detecting disease-related genes with higher power than  each platform individually.} \cite{yang2017cancer} collectively model multi-omics platforms with an unsupervised Bayesian nonparametric framework to identify driver genes. An integrative approach for biomarker discovery using latent components coupled with supervised analysis and prediction has been implemented by \cite{singh2019diablo}. \cite{maity2020bayesian}, on the other hand, predicted patients' prognosis from their molecular profiles with a Bayesian structural equation modeling.

{The fourth group of studies concerns integrative clustering of subjects to detect biologically relevant subpopulations. These approaches typically model latent components or specify similarity scores among subjects to detect clusters of patients  potentially corresponding to disease subtypes. } \cite{lock2013bayesian} proposed Bayesian consensus clustering (BCC), which gives an overall clustering of data sources, also allows source-specific clustering. However, misspecifying the number of clusters could easily lead to under or over fitting. Dependence of each platform to the overall clustering is always difficult to justify as well, for the reason that the parameters which control the adherence of platform to the overall clustering depend on number of clusters; the strength of adherence is therefore hard to specify or learn.  {The iCluster methods \citep{mo2018fully, mo2013pattern, shen2009integrative} used orthogonal latent variables to represent complex molecular structures,  achieving dimension reduction and revealing tumor subtypes. Many other integrative clustering techniques  focus on optimizing survival curves separation while simultaneously detecting  patients groups with similar molecular traits \citep{ahmad2017towards, coretto2018robust}. \cite{park2021integrating} recently proposed a kernel-learning method that utilizes the spectral clustering framework with a flexible choice of similarity measure to detect cancer patient subtypes. }




\section{Approach}
The central hypothesis behind our proposed approach is that multiple omics platforms provide contrasting, yet intrinsically interrelated, information about the underlying molecular drivers of  phenotypes or clinical outcomes. Consequently, to improve tailored therapy and monitoring for cancer patients, it is  crucial to aggregate information from the  platforms  to uncover novel patient subpopulations, e.g., cancer subtypes, and quantify the associations of the somatic measurements with the study endpoints. The challenges posed by these integrative analyses include not only high dimensionality, but also substantial probe-to-probe dependencies caused by complex biomarker  interactions through  gene or metabolic pathways. 

To tackle these challenges, we propose a flexible  Bayesian nonparametric framework for analyzing multi-domain, complexly structured, and high throughput modern array data and detecting different  biological structures in different platforms. The model that can be  summarized as follows. We utilize the flexibility of    Poisson-Dirichlet processes (PDPs) to discover the  biological dependence of the multivariate vectors associated with the biomarkers, both within and between the omics domains. The PDPs naturally facilitate dimension reduction in the large number of omics biomarkers,  allowing  variable selection in these high-dimensional settings, and providing a catalogue of genomic aberrations associated with the cancer-related clinical outcomes. A nested univariate Dirichlet process induces an overall, global clustering of the patients across the disparate data sources to detect an unknown number of tumor subtypes.  The MCMC procedure for fitting the model is computationally feasible due to the application of data-squashing techniques developed in \cite{guha2010posterior} that drastically reduce the computational times for high-throughput omics datasets.

The rest of the paper is organized as follows.  Section \ref{section: MEI} introduces the details of the integrative Bayesian hierarchical model in three major parts. Specifically, Section \ref{subection: BiClstr} introduces a novel bidirectional clustering technique that achieves nonparametric dimension reduction and platform-specific detection of unknown pathways. In Section \ref{subsection: Linkbi}, we propose a method to account for the similarity of the elements of platform within each unsupervised bidirectional cluster.  Section \ref{subsection: Extrgen} discusses a similar hierarchical construction of \cite{Teh04hierarchicaldirichlet} that allows the borrowing of strength within and across platforms. Section \ref{section: VS} describes a variable selection procedure that can effectively filter useful information about any available clinical outcome to improve the quality and stability of the subtypes discovery. Posterior inference procedures are illustrated in Section \ref{section: PI}. Through simulations in Section \ref{subsection: SIM}, we demonstrate the accuracy of the clustering mechanism using simulated datasets. In Section \ref{subsection: TCGA}, we analyze the motivating gene expression and DNA methylation datasets for lung cancer. We conclude with some remarks in Section \ref{section: DIS}. 

\section{Methods}
\subsection{An Integrative Bayesian Hierarchical Model for Multiple Omics~Domains}\label{section: MEI}
As the same patient tumor samples are profiled using different assays, the data are inherently dependent across the various platforms. 
We develop a statistical framework for the joint modeling of discrete and continuous somatic measurements  arising from multiple omics domains.  
We begin with some basic notation. Assume that we wish to analyze $T \ge 2$ platforms or omics datasets for a total of $n$ matched patients or tissue samples for a certain disease. 
In general, the observations for a given platform may be continuous (e.g., gene expression data), count (e.g., copy number variation data), or proportion  (e.g., DNA methylation data).

\begin{figure}[!tpb]
\centering
	\includegraphics[width=7.5cm,height=3cm]{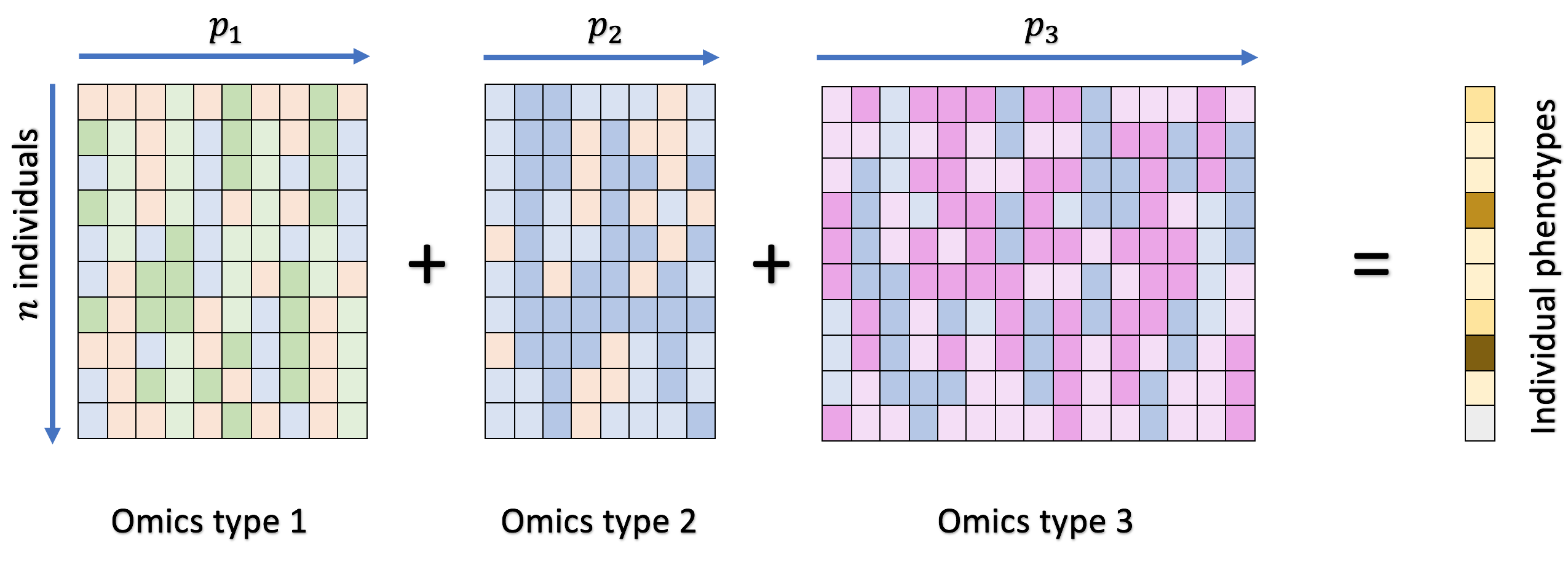}
	\caption{A graphical illustration of data structure: $n$ individuals measured at $T=3$ platforms whose survival time and censoring indicator are recorded as clinical outcomes.}
	\label{figure: datastructure}
\end{figure}

Let $\bm{X}_t$ represent the matrix of genomic or epigenomic measurements associated with the $t$th platform.  The rows of matrix $\bm{X}_t$ represent the $n$  patients, and the matrix columns represent the $p_{t}$ biomarkers. 
Although $n$ itself is fairly large, it  is typically far smaller than the number of biomarkers in each dataset, so that  $n \ll \min_{t} p_t$. A generic element of matrix $\bm{X}_t$ is  $X_{ijt}$ for row (patient) $i=1,2,\dots,n$, and column (probe) $j=1,2,\dots,p_{t}$. 
Patient phenotypes, {such as overall survival times, are denoted by $w_{i}$ for $i = 1, \cdots, n$. See Figure \ref{figure: datastructure} for a graphical representation. With $t_{i}$ denoting the failure time and $C_i$ denoting the independent censoring time, the observed survival time $w_i=\min\{t_i, C_i\}$ and censoring indicator $\Delta_i = \mathcal{I}(t_{i} \le C_i)$.}


\subsubsection{Bidirectional clusters:  dimension reduction and platform-specific detection of biomarker  dependencies}\label{subection: BiClstr}

Since the number of patients or rows, $n$, is  far smaller than the large number of biomarkers or columns, $p_t$,  of matrix $\bm{X}_t$, collinearity is a common phenomenon. This so-called ``large $n$, larger $p_t$'' collinearity issue is known to result in instability with traditional statistical  techniques. Apart from the purely mathematical reason that $n \ll p_t$, collinearity also occurs because  unknown sets of biomarkers are expected to be involved in unknown or partially biological processes (e.g., pathways), some of which are involved in  the development and progression of   diseases such as cancer. This results in co-expression and, therefore, high pairwise sample correlations of those sets of biomarker-specific column vectors of length~$n$.

 As a first  step, we apply a platform-specific transformation, $z_t(\cdot)$,  {that maps the support of continuous somatic measurements into the unbounded real line. Subsequently, the data is   modeled nonparametrically to flexibly capture its distinctive characteristics. 
 For continuous gene expression data,  the identity transformation, $z_t(x)=x$ is  appropriate. An appropriate transformation for proportions, e.g., DNA methylation data, is the logit function, $z_t(x)=\log(x/(1-x))$. This results in a transformed data matrix denoted by $\bm{Z}_t$ for platform $t$, with generic element, $Z_{ijt}  = z_t\bigl(X_{ijt}\bigr) \in \mathcal{R}$, which is then modeled by a Bayesian nonparametric model.} Let the column vectors of matrix  $\bm{Z}_t$ of length $n$ each  be denoted by $\bm{z}_{1t},\ldots,\bm{z}_{p_tt}$. 
We rely on  model-based, bidirectional   clustering to mitigate the issues of    collinearity and high-dimensionality of matrix $\bm{Z}_t$. Clustering is performed  along both dimensions of the platform-specific matrices.

\paragraph{Platform-specific  column clusters} \quad The $p_t$  biomarkers or probes of matrix $\bm{Z}_t$ are assumed to belong to an unknown number, $K_{t}$, of latent column clusters, with each latent column cluster consisting of  probes with highly similar $n$--variate column vectors, $\bm{z}_{jt}$.
The  cluster membership information is represented by a  {\it column  allocation variable}, $c_{jt}$, taking values in the set $\{1,2,\cdots,K_{t}\}$ and identifying the  cluster to which  the $j$th probe  belongs. 
 That is, for $k=1,2,\cdots,K_{t}$, the  event $[c_{jt}=k]$ implies that the $j$th probe  belongs to the $k$th latent column cluster in the $t$th platform. Given  the column allocation variables, as the latent clusters consist of similar probes,  the statistical analysis can performed on the level of the smaller number of clusters rather than the large number of individual probes of the $t$th platform. 
 This strategy  downsizes the original {\it large $n$, larger $p_t$} problem to a more manageable  {\it large $n$, small $K_{t}$} problem for which  more stable inferences are achieved.
 
  For a stochastic model for the probe-to-column-cluster
allocations, we rely on the
partitions induced by the two-parameter Poisson-Dirichlet process:
$
c_{1t}, c_{2t}, \ldots, c_{p_t t} \sim PDP(\alpha_1,  d_{t}), 
$
with platform-specific discount parameter $0\le d_{t}<1$ and mass parameter $\alpha_1>0$.  Subsequently, this partitioning of the biomarkers  allows  the experimental validation and discovery of unknown biological processes identified by the latent clusters. The PDP was originally formulated  by \cite{perman1992size}, and further developed by \cite{pitman1995exchangeable} and \cite{pitman1997two}. \cite{lijoi2010models} reviews Gibbs-type Bayesian nonparametric priors,   such as Dirichlet processes and PDPs. Previous works that have utilized the sparsity induced by  PDPs and Dirichlet processes to achieve dimension reduction include \cite{medvedovic2004bayesian}, \cite{kim2006variable},
 \cite{lijoi2007bayesian},  \cite{dunson2008bayesian},  \cite{dunson2008kernel}, and \cite{guha2016nonparametric}.

  Since the cluster labels are arbitrary, we may assume without loss of generality that $c_{1t}=1$, i.e., the first probe of  platform $t$  is assigned to the first platform-specific column cluster. 
 For a subsequent biomarker indexed by $j>1$, suppose there are $K_{jt}$ distinct clusters among the first $j$ allocation variables, $c_{1t}, \cdots, c_{jt}$, with the $k$th column cluster containing $n_{jkt}$ number of probes. Then the predictive prior probability that the $(j+1)$th probe belongs to the $k$th column cluster is as follows: $P\bigl(c_{j+1,t}=k \mid c_{1t},\cdots,c_{jt}\bigr)$ is proportional to $(n_{jkt} -d_{t})$
for column cluster $k=1,\cdots,K_{jt}$, and proportional to 	$(\alpha_1+K_{jt} d_{t})$ when $k=(K_{jt}+1)$,
with the latter option representing the event that $(j+1)$th probe opens a new latent column cluster.  
Despite the sequential description, the $p_t$ allocation variables are exchangeable in the prior \citep{lijoi2010models}. Obviously, $K_{t}$, the total number of column clusters among the $p_t$ probes, is  equal to random variable $K_{p_tt}$  of the above sequential description. Larger  (near 1) values of 
 PDP discount parameter $d_{t}$ imply a higher propensities of the probes to open new latent column clusters, and therefore, larger $K_{t}$.

Within each platform, the PDP reduces the effective dimension  of the probes because the unknown number of latent clusters, $K_t$, although random,  is much smaller than the number of probes, $p_t$. This is  because as $p_t$ grows, random variable $K_{t}$  is asymptotically equivalent to  $\alpha_1  \log p_t$ if PDP discount parameter $d_{t}=0$, and to $T_{d_{t}, \alpha_1}  p_t^d$ if $0 < d_{t} < 1$, for  some positive random variable $T_{d_{t}, \alpha_1}$ \citep{lijoi2010models}.
  Thus, irrespective of the value of $d_{t}$, the number of latent column clusters is of a smaller order of magnitude than the number of probes, resulting in dimension reduction. Because it results in a logarithmic rather than polynomial reduction in the number of probes, the number of  Dirichlet process clusters (i.e., when $d_{t}=0$) is asymptotically of a smaller order than the number of non-Dirichlet   PDP  clusters. The PDP discount parameter is given the mixture prior $\frac{1}{2}\delta_0+\frac{1}{2}U(0,1)$, where $\delta_0$ denotes the point mass at 0, and represents a Dirichlet process with mass parameter $\alpha_1$. We make posterior inferences on  PDP discount parameter $d_t$ to find the omics  platform-specific column allocation pattern that best matches the data.

 \paragraph{Global (platform-independent) row clusters} \quad  In many investigations, a primary goal of 
 integrative genomic, epigenomic, and transcriptomic analysis is the discovery of unknown disease subtypes. The following model feature is ideally suited for  this purpose. 
Conditional on the column clusters of the $T$ platforms, and {\it irrespective of platform}, we suppose that the $n$ rows of the matrices $\bm{Z}_1,\ldots,\bm{Z}_T$ are  mapped to a much smaller, unknown number, $H$, of latent row or patient clusters. Intuitively, each latent cluster represents a subpopulation of patients whose $K_{t}$--variate compressed row vectors in matrix $\bm{Z}_t$ are highly similar across   platforms.  The global row clusters  partition the $n$ individuals with the disease into  $H$  latent subpopulations on the basis of their omics profiles. These subpopulations may be indicative of (potentially unknown) disease subtypes, as we shall later demonstrate for the  TCGA lung cancer data.

We denote the subject-to-row-cluster  mappings by $r_{1},\cdots, r_{n}$, with each row allocation variable taking values in $\{1, 2,\cdots, H\}$. The number of global row clusters,  $H$, is random, and equal to the number of partitions induced on the  patients by a Dirichlet process with mass parameter $\alpha_2$. 
 This modeling strategy further distills the already reduced  {\it large $n$, small $K_{t}$} problem to an even more manageable {\it small $H$, small $K_{t}$} problem,  further facilitating reliable posterior inferences.

\subsubsection{Linking    bidirectional clusters to  somatic measurements} \label{subsection: Linkbi}

The next level of the hierarchical model
  accounts for  the  similarity of the  matrix $\bm{Z}_t$ elements in each unsupervised bidirectional cluster. Conditional on  the bidirectional cluster allocations, we envision the existence of a reduced-dimensional version of matrix $\bm{Z}_t$ as an order $H \times K_{t}$ {\bf latent  matrix}, $\bm{\Phi}_{t}=((\varphi_{hkt}))$, for row cluster $h=1,\ldots,H$ and platform-specific column cluster $k=1,\ldots,K_{t}$. Each element of the lower-dimensional  matrix $\bm{\Phi}_{t}$
 has a one-to-one mapping with a bidirectional cluster, and it assimilates information from the multiple elements of transformed data matrix $\bm{Z}_t$  mapped to that cluster. Further details are provided below.
 
 \paragraph{Likelihood function} \quad The transformed somatic measurements belonging to a bidirectional cluster are modeled as the cluster-specific latent matrix elements plus white Gaussian noise. This construction induces  high correlation between    $\bm{Z}_t$ elements belonging to the same bidirectional cluster.
 {More formally, suppose subject $i$ belongs to the $h$th global row cluster, and probe $j$ in matrix  $\bm{Z}_t$ belongs to the $k$th platform-specific column cluster, i.e.,  $r_{i}=h$ and  $c_{jt}=k$.   Then   $Z_{ijt}$ is related to its mapped latent matrix element, $\varphi_{hkt}$, as $	Z_{ijt}  \sim$ $N(\varphi_{hkt},\sigma^2)$. For subject $i=1,\ldots,n,$, probe $j=1,\ldots,p_t$, and platform $t=1,\ldots,T$,  we obtain $Z_{ijt}  \overset{\text{indep}}{\sim}$ $N(\varphi_{r_{i}c_{jt}t},\sigma^2)$.}


\paragraph{Latent matrix elements}  \quad 
 To achieve  a flexible and  sparse model,  we allow the elements of  latent matrix $\bm{\Phi}_{t}$ to  mutually communicate and borrow strength:
 $\varphi_{hkt} \stackrel{iid}\sim G_{t}$ for global row cluster $h=1,\ldots,H,$ and  platform-specific column cluster $k=1,\ldots,K_{t}$.
The set of unknown  univariate distributions, $\{G_{t}:t=1,\ldots,T\}$,  are themselves given  a common, nonparametric  Dirichlet process prior on the space of all univariate distributions:  $G_{t} \stackrel{iid}\sim DP(\alpha_3, G_0)$ for platform $t=1,\ldots,T$, 
   mass parameter $\alpha_3>0$ and  a random univariate base distribution,  $G_0$.

The nonparametric nature of the  Dirichlet process  allows the model to flexibly capture the features of  data matrix $\bm{Z}_t$, e.g., mean-variance relationships \citep{lijoi2010models}. Furthermore, being a realization of a Dirichlet process, each distribution $G_{t}$ is discrete. This causes a substantial proportion of the latent matrix elements to be tied.
In fact, the  number of distinct values in latent matrix $\bm{\Phi}_{t}$ is relatively small and  is  asymptotically equivalent to $\alpha_3 (\log n  + \log p_t)$. Thus,   dimension reduction is an integral feature of the model, allowing  scalable  inferences for the large the number of subjects and probes grows in next-generation sequencing genomic and epigenomic datasets.

\subsubsection{Extraction of genomic information from multiple domains} \label{subsection: Extrgen}

\begin{figure}[!tpb]
\centering
	\includegraphics[width=7.5cm,height=4.5cm]{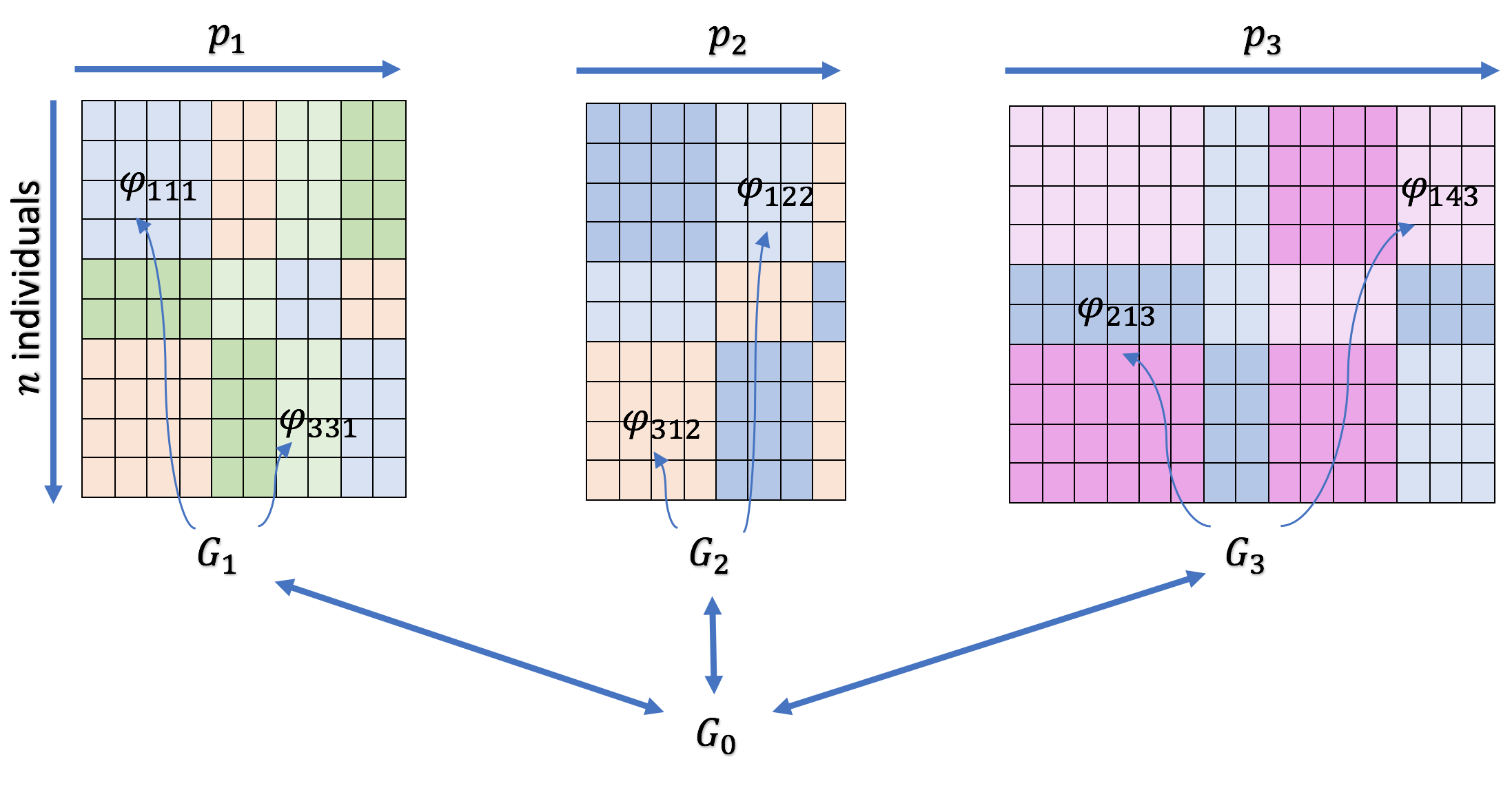}
	\caption{A graphical representation of bidirectional clustering and a hierarchical construction of Dirichlet process. Communications among the various disease subtype--platform combinations is made possible by a higher-level Dirichlet process specification for the base distribution $G_0$.}
	\label{figure: model}
\end{figure}

In multiomic datasets,
biological information about a disease  can be extracted  from latent  biomarker pathways  and patient subpopulations. To capture this aspect of the data, similar to the hierarchical construction of \cite{Teh04hierarchicaldirichlet}, we model the random base distribution $G_0$ of the Dirichlet process prior to allow the bidirectional clusters of the various disease subtype--platform combinations to mutually communicate. This is achieved by a higher-level Dirichlet process specification for base distribution,  $G_0$, itself: $G_{0}\sim DP\bigl(\alpha_4, N(\mu_0,\tau_0^2)\bigr)$.
The discreteness of distribution $G_0$ due to the   Dirichlet process \citep{lijoi2010models} is critically important because  it causes the  atoms of  the discrete distributions, $\{G_{t}\}$, to be shared across disease subtype and platform. Consequently, several latent matrix elements of a disease subtype--platform combination are  identical to  others belonging to a different  disease subtype--platform combination. In this manner,  we combine  unsupervised bidirectional clustering with an effective use of Bayesian mixture~models  to achieve data integration through sharing of information. See Figure \ref{figure: model} for a graphical representation of the hierarchical structure.

\subsection{Models for Survival Outcomes and Gene Selection}\label{section: VS}

We expect only a small fraction of the probes or biomarkers to be associated with the clinical outcome of interest. Extracting  the relevant biological features  and their relationships with  patient  outcomes  reveals important and possibly unknown biological mechanisms of   disease initiation and progression that can  be subsequently validated in a wet lab. 
An important part of effective feature extraction is the elimination of redundancy in the lower-dimensional latent structure detected by the  hierarchical model of Section \ref{section: MEI}. 

To achieve this, we regard the platform-specific latent clusters as potential predictors and  
discard repeats in these  predictors, so that  the number of possible predictors is smaller than the number of total clusters. 
As mentioned, there are $K_{t}$ column clusters in the $t$th dataset. We  aggregate the latent clusters over the $T$ platforms $\bm{Z}_t$ by merging clusters that share the same  columns of their latent matrices. This reduces the total (across-platform) number of  column clusters  to $K\le \sum_{t=1}^{T}K_{t}$. After this merging procedure, let the number of   probes associated with the $k$th column cluster be denoted by $N_{k}$,  where $k=1,\ldots,K$. 

Next, we imagine that each merged column cluster elects from among its member probes a representative, denoted by $\bm{\mu}_k$. The column cluster representatives are selected with a priori equal probability from the $N_{k}$ covariates. {However, the \textit{posterior} probabilities of the column cluster representative are usually different for the   probes.}  
We rely on an additive regression model  that  accommodates potential nonlinear functional relationships to detect a typically small subset of the $K$  cluster representatives as the regression predictors. By applying extensions of the spike-and-slab approaches \citep{george1993variable, karpenko2010relational, brown1998multivariate}, the additive regression model is  	$y_i\overset{\text{indep}}{\sim} N(\eta_i,\tau^2)$, where 
\vspace{-.1 in}
\begin{equation}
\eta_i=\beta_{0} + \sum_{k=1}^{K}\gamma_{1k} \beta_{1k}\mu_{ik} + \sum_{k=1}^{K}\gamma_{2k}h(\mu_{ik},\beta_{2k}), \label{reg}
\end{equation}
 $h$ is a nonlinear function and {the $\beta$'s are regression coefficients. For an AFT survival outcomes model, we assume $y_i = \log(t_i)$ is a Gaussian regression outcome that is transformed from the uncensored failure time $t_{i}$. As usual, observed survival time $w_i=\min\{t_i, C_i\}$  for censoring time $C_i$.} For  column cluster $k$, expression (\ref{reg}) implicitly relies on a  triplet of indicators, $\bm{\gamma}_k=(\gamma_{0k}, \gamma_{1k},  \gamma_{2k})$, of which exactly one indicator equals $1$ and the other two indicators equal $0$. For example, if $\gamma_{0k}=1$, then none of the $N_k$ probes belonging to column cluster $k$ are associated with the responses. On the other hand, if $\gamma_{1k}=1$, then  cluster representative $\bm{\mu}_k$ appears as a simple linear regressor. Lastly, if $\gamma_{2k}=1$, then $\bm{\mu}_k$ is a nonlinear regressor  in  (\ref{reg}). The total number of linear predictors, nonlinear predictors, and non-predictors in expression (\ref{reg}) are then $k_1=\sum_{k=1}^{K}\gamma_{1k}$, $k_2=\sum_{k=1}^{K}\gamma_{2k}$, and $k_0=K - k_1 - k_2$, respectively. We collectively denote the $K$ column cluster-specific triplets of indicators by $\bm{\gamma}=(\bm{\gamma}_1,\cdots, \bm{\gamma}_{K})$.

Due to their interpretability and  computational efficiency, one could choose the nonlinear function as order-$s$ splines with $v$ number of knots \citep{de1978practical, hastie1987generalized, denison1998automatic}. Other  options include reproducible kernel Hilbert spaces, nonlinear basis smoothing splines, and wavelets \citep{mallick2005bayesian, eubank1999nonparametric}. In general,
for nonlinear functions $h$ having a linear representation, 
let  $\bm{U}_{\bm{\gamma}}$ be  the matrix of $n$ rows  consisting of the intercept and  $(k_1+k_2)$ regression predictors. Specifically, if we use order-$s$ splines with $v$ number of knots in equation (\ref{reg}), then the number of columns of $\bm{U}_{\bm{\gamma}}$ is  $\text{col}(\bm{U}_{\bm{\gamma}})=k_1 + k_2(v+s) +1$. 

With $[\cdot]$ denoting densities of random variables, we specify the prior of indicators $\bm{\gamma}$ as $[\bm{\gamma}] \propto \omega^{k_{0}}_0\omega^{k_{1}}_{1}\omega^{k_{2}}_2\cdot \mathcal{I}\big(\text{col}(\bm{U}_{\bm{\gamma}})<n\big)$,
where probability vector $\bm{\omega}=(\omega_0, \omega_1, \omega_2)$ is given a Dirichlet distribution prior: $\bm{\omega}\sim \mathcal{D}_3(1,1,1)$.  The truncated prior for $\bm{\gamma}$ ensures model sparsity and prevents overfitting. 
Finally, conditional on the parameter $\tau^2$, we assign a   $g$ prior \citep{zellner1986assessing}   to the regression coefficients:
$\bm{\beta}_{\bm{\gamma}}|\bm{\Sigma} \sim N\big(\bm{0}, \sigma^2_{\beta}\tau^{-2}(\bm{U}_{\bm{\gamma}}' \bm{U}_{\bm{\gamma}})^{-1}\big)$.

\subsection{Posterior Inference}\label{section: PI}
 The model parameters are initialized by a naive, separate analysis of each omics platform. Thereafter, the parameters are iteratively updated by MCMC methods.  Due to the complex nature of the posterior inference, the procedure is carried out in two  stages: (1) platform-specific column cluster  and global row cluster detection, followed by (2) predictor discovery for each platform:

\begin{enumerate}

	\item[]\textbf{Stage 1} \quad 
	Focusing only on the multiomic  measurements, the column allocation variables, $c_{jt}$, subject--to--row-cluster allocation variables $r_{i}$, and the latent matrix elements $\varphi_{hkt}$, defined in \ref{section: MEI}, are iteratively updated. More specifically:
	
	\begin{itemize}
   \item[]\textbf{Stage 1a} \quad   Using an unrestricted MCMC run, the posterior probability of pairwise clustering of probes are first computed by an MCMC sample. Applying the technique of \cite{dahl2006model}, these pairwise probabilities are used to obtain a point estimate for the column cluster allocation variables, $c_{jt}$, for $j=1,\ldots,p_t$, $t=1,\ldots,T$, called the {\it column least-squares allocation}.
   
   \item[]\textbf{Stage 1b} \quad  Conditional on the column least-squares allocation, a second MCMC sample is generated. Again using the technique of~\cite{dahl2006model}, we compute a point estimate, called the  {\it row least-squares allocation}, for the subject--to--row-cluster allocation variables $r_1,\ldots,r_n$.  
   
	\item[]\textbf{Stage 1b} \quad  Conditional on the bidirectional  allocation variables, a third MCMC sample is generated to estimate the latent matrix elements, $\varphi_{hkt}$. 
   
	\end{itemize}

\item[]\textbf{Stage 2} \quad Conditional on the  column and global  row least-squares allocations and latent matrix elements, the patient-specific responses are  analyzed using a fourth MCMC sample. In this way, we infer the column cluster predictors and cluster representatives, defined in Section \ref{section: VS}, and associated with the disease outcomes. 


\end{enumerate}

\paragraph{Predictor detection via Bayesian FDR control} \quad When the number of merged clusters $K$ is large, we  control the expected Bayesian FDR at nominal level $\alpha$ \citep{newton2004detecting,muller2006fdr,morris2008bayesian}. Let  
$b_k$ denote the posterior probability that the $k$th merged  cluster is a predictor,   $k=1,\ldots,K$. 
Then an MCMC estimate of $b_k$ is 
$\hat{b}_k = \frac{1}{M}\sum_{m=1}^{M}\mathcal{I}\bigl(\gamma_{1k}^{(m)} + \gamma_{2k}^{(m)} =1\bigr)$, where $M$ is the number of MCMC  samples. We  sort $\hat{b}_1,\ldots,\hat{b}_K$ in descending order to obtain  $\hat{b}_{(1)}\ge\ldots\ge\hat{b}_{(K)}$. For a cutoff $\psi_\alpha = \hat{b}_{(l)}$, where $l = \max\{k^*: \sum_{k=1}^{k^*}(1 - \hat{b}_{(k)}) \le \alpha\}$. Then the set of merged clusters $\{k: \hat{b}_k > \psi_\alpha \}$ are declared to be predictive of the clinical outcomes.  


\section{Results}\label{section: NS}
  \subsection{Simulation}\label{subsection: SIM}

We investigated the  clustering accuracy of the proposed methodology using artificial datasets for which the column clusters and global row clusters   are known. {No cluster predictors are generated for the responses in this simulation.} The data generation procedure consisted of the following two parts:

  \paragraph{Probe generation}  We generated a dataset consisting of $n=70$ subjects and $T=2$ platforms,  with each platform containing $250$ probes, i.e., $p_1=p_2=250$.  Probes for each data are generated from a discrete measure of disease subtype--platform combinations convolved with Gaussian noise, which is further generated from another  discrete base distribution. The global row clusters are generated by a finite mixture. We generated the following quantities to obtain transformed data matrix $\bm{Z}_{t}$ of dimension $70 \times 250$. 

    \begin{enumerate}[\quad (1)]

  \item True subject--to--row-cluster  mappings: Given $n=70$, generate global row clusters labels $\tilde{r}_{1},\cdots, \tilde{r}_{n}$ from a multinomial distribution with equal probabilities $(\frac{1}{3}, \frac{1}{3}, \frac{1}{3})$. Thus, $\tilde{r}_{i}\in\{1, 2, 3\}$.

  \item True column allocation variables: For platform $t=1, 2$,  generate $\tilde{c}_{1t}, \cdots, \tilde{c}_{p_{t}t}$ as the partitions induced by a PDP with true discount parameters $\tilde{d}_{11} = 0.2$ and $\tilde{d}_{12} = 0.25$, and with common mass parameter $\tilde{\alpha}_1=10$. The true number of clusters, $\tilde{Q}_t$, was thereby computed for this non-Dirichlet allocation.

  \item True base distribution $\tilde{G}_0$: $\tilde{G}_0$ is generated from $DP(\tilde{\alpha}_4, N(\tilde{\mu}_0, \tilde{\tau}_0^2))$ according to the stick-breaking process of \cite{sethuraman1982convergence} and \cite{sethuraman1994constructive} with $\tilde{\alpha}_4=10$, $\tilde{\mu}_0=0$, and $\tilde{\tau}_0=1$.

\item True discrete measure of disease subtype--platform combinations $\tilde{G}_{1t}$: $\tilde{G}_{1t}$ is generated from $DP(\tilde{\alpha}_3, \tilde{G}_{0}$) with $\tilde{\alpha}_3=10$, again according to a stick-breaking process.

\item Probes for dataset $t$: We generate the latent matrix elements, $\tilde{\varphi}_{hkt}$, as i.i.d samples from $\tilde{G}_{1t}$. The true mapping between $\tilde{\varphi}_{hkt}$ and $Z_{ijt}$ are $\tilde{r}_{1},\cdots, \tilde{r}_{n}$ and $\tilde{c}_{1t}, \cdots, \tilde{c}_{p_{t}t}$, and the true noise parameter is $\tilde{\sigma}=0.2$. The elements of a data matrix $\bm{Z}_{t}$ are then distributed as 
$ Z_{ijt} \mid \tilde{r}_{i}=h,  \tilde{c}_{jt}=k \overset{indep}{\sim}N(\tilde{\varphi}_{hkt},\tilde{\sigma}^2)$.
\end{enumerate}

    \paragraph{Unsupervised subtypes discovery} Assuming an AFT survival model, apply the procedure as described in Section \ref{section: VS} with linear splines and $v=1$ knot per spline. Selecting one probe from each cluster as the representative, variable selection was performed independently for these two data types. Posterior inferences were made according to the stages described in Section \ref{section: PI}. We evaluate the column clustering and global row clustering accuracy based on the post-processed MCMC samples obtained in Stages 1 and 3 of Section \ref{section: PI}. 

    Applying the technique of \cite{dahl2006model} developed for Dirichlet process models, point estimates for the column allocations $\hat{c}_{1t}, \cdots, \hat{c}_{p_{t}t}$ are computed. For the full set of probes from the two data types, we estimated the accuracy of the column allocation by the estimated proportion of correctly clustered probe pairs, defined as
$
	\hat{\varkappa}_{t}=\frac{1}{\binom{p_{t}}{2}}\underset{j_1\neq j_2\in (1,\cdots,p_{t})}{\sum}\mathcal{I}\big( \mathcal{I}(\hat{c}_{j_{1}t} = \hat{c}_{j_{2}t})= \mathcal{I}(\tilde{c}_{j_{1}t} = \tilde{c}_{j_{2}t})\big)$.

Also applying the technique of \cite{dahl2006model}, point estimates for subject--to--row-cluster  mappings $\hat{r}_{1},\cdots, \hat{r}_{n}$ were computed. We estimated the accuracy of subject--to--row-cluster  mappings in the same way:
$
	\hat{\vartheta}=\frac{1}{\binom{n}{2}}\underset{i_1\neq i_2\in (1,\cdots,n)}{\sum}\mathcal{I}\big( \mathcal{I}(\hat{r}_{i_{1}} = \hat{r}_{i_{2}})= \mathcal{I}(\tilde{r}_{i_{1}} = \tilde{r}_{i_{2}})\big)$. 
High values of $\hat{\varkappa}_{t}$ and $\hat{\vartheta}$ are indicative of high clustering accuracy for all $p_{t}$ probes and all $n$ subjects.

 Table \ref{table:column_percent} displays the mean percentages $\hat{\varkappa}_{t}$ and estimated $R^2$ averaged over $50$ independent replications. The mean proportion of correctly clustered subject pairs $\hat{\vartheta}$ is 0.955 (0.042), with standard error shown in parentheses. We find that significantly less than 5\% pairs of subjects are incorrectly clustered  in the same global row clusters out of $\binom{70}{2}=2415$ different subjects pairs, and significantly less than 3\% pairs of probes are incorrectly clustered  among $\binom{250}{2}=31125$ different probe pairs in each platform. Thus, over $50$ replication, we observed highly accurate clustering-related inferences for the global row clusters as well as for the full set of $p=250$ probes of every platform. The high accuracy is also reflected by the estimated $R^2$'s, all of which are significantly greater than $90\%$.

	
To further demonstrate the performance of our method, we generated $12$ setups, with $T=2$ data types. We generated $h=3, 4, 5$  global row clusters using the same way as the previous simulation in Section \ref{subsection: SIM}. For each value of $h$,  we varied the data noise $\tilde{\sigma}$ among $0.2, 0.3, 0.4, 0.5$. For each set up, we replicated the procedure $50$ times. Therefore, we covered situations ranging from few global row clusters with small noise to many global row clusters with large noise.

\begin{figure}[!pb]
\centering
	\includegraphics[width=80mm,height=5cm]{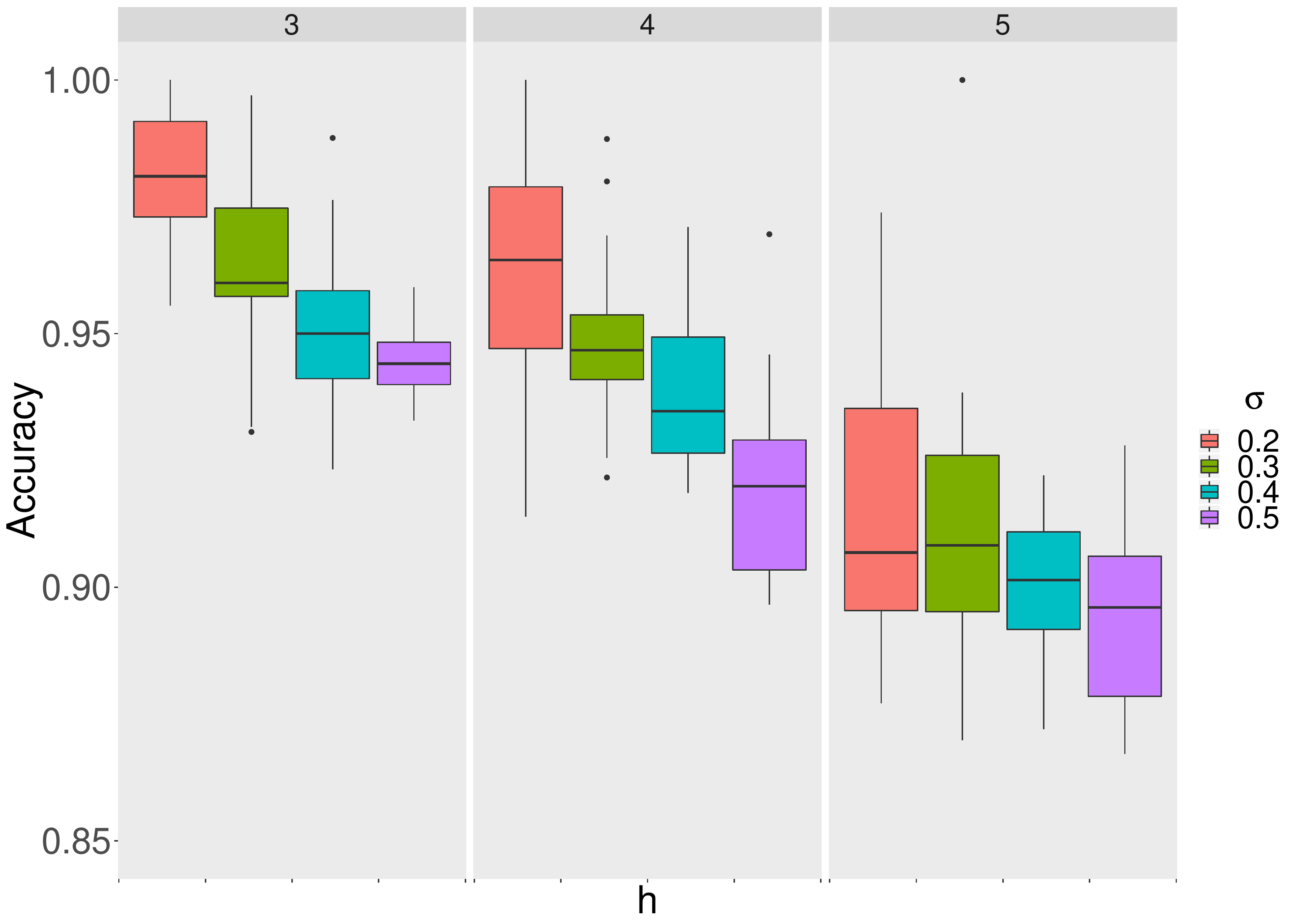}
	\caption[0.8\textwidth]{Side-by-side boxplots comparing combination of different numbers of global row clusters and $\sigma$ values. }
	\label{figure: row}
\end{figure}

The subject--to--row-cluster  mappings accuracy is consistently going up as $h$ reaches 5 as shown in Figure \ref{figure: row}. It moderately goes down as more noise involved; however, it remains above $90\%$ when global row clusters is $5$. 

\begin{figure}[!tpb]
\centering
	\includegraphics[width=86mm]{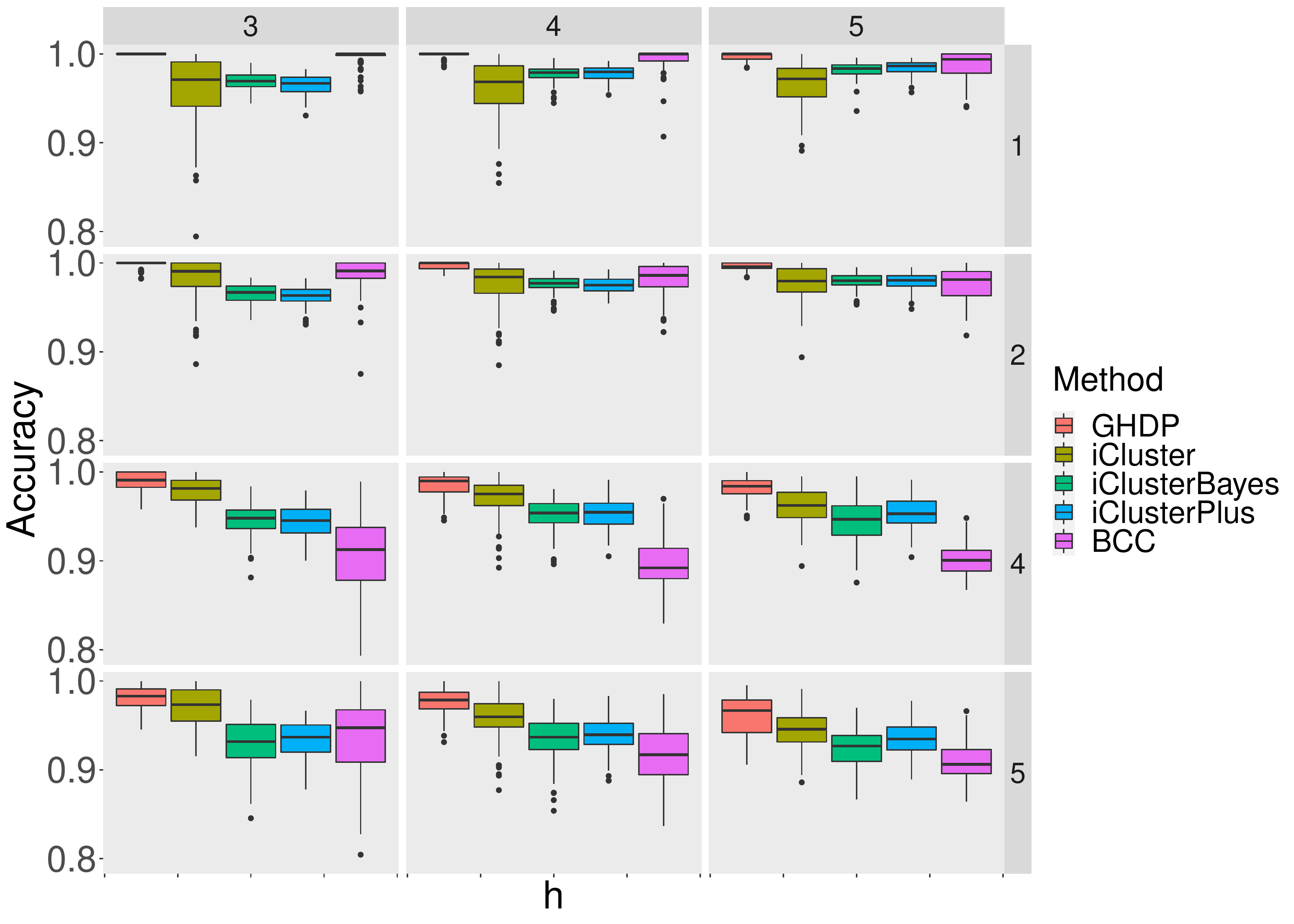}
	\caption[0.8\textwidth]{Side-by-side boxplots comparing the proposed method with iCluster, iClusterBayes, iClusterPlus, and BCC over $12$ combinations of different numbers of global row clusters and noise levels. }
	\label{figure: BCC}
\end{figure}

{Furthermore, we evaluated our model's performance in subject--to--row-cluster  mappings  by comparing it with other model-based integrative clustering methods. These methods include iCluster, iClusterPlus, iClusterBayes \citep{shen2009integrative, mo2013pattern, mo2018fully}, and Bayesian consensus clustering (BCC) \citep{lock2013bayesian}. For a fair comparison, we simulated data as a mixture of multivariate Gaussians with non-trivial correlation structure and varying noise levels using the MixSim R-package \citep{melnykov2012mixsim}. In specific, the package is used to simulate $h = 3, 4, 5$ cluster-specific mean vectors, $p_{t} \times p_{t}$ cluster specific full covariance matrices for $p_{t} = 250$, $t = 1, 2$, and with different noise levels. We repeated the simulation settings $100$ times and compared the results with other competing methods.} For BCC, we assume that the data sources are equally adhering to the overall clustering, and specify the true number of global row clusters as the number of overall clustering. Figure \ref{figure: BCC} depicts  boxplots of the estimated proportion of correctly clustered subject pairs for our method and others. Our method remains above $95\%$ subject--to--row-cluster allocation accuracy for all $12$ set ups, with considerably smaller standard errors in most set ups.

\subsection{An application to TCGA datasets}\label{subsection: TCGA}
The proposed methodology is applied to a  lung cancer  dataset available from TCGA. TCGA includes tumor samples from more than $500$ patients with lung cancer. Analyzing different platforms illuminates some of the biomolecular characteristics in lung cancer. For example, the iCluster methods \citep{mo2018fully, mo2013pattern, shen2009integrative} developed a joint latent variable model for integrative clustering of copy number and gene expression data. \cite{kim2015integrative} introduced an integrative phenotyping framework (iPF) for disease subtype discovery. {LUSC and LUAD are two main subtypes of lung cancer. They are the most common type of NSCLC, and LUAD has a high rate of mutations compared to other cancers. Previous studies regarding molecular profiling of LUSC and LUAD have been extensively discussed \citep{carrot2020comprehensive, cancer2014comprehensive, cancer2012comprehensive}. Four subtypes of LUSC were molecularly described, classical, basal, secretory, and primitive. Three distinct molecular subtypes of LUAD have been discovered based on gene expression data, proximal inflammatory (PI), proximal proliferative (PP), and terminal respiratory unit (TRU). These subtypes were observed to be relevant to pathways and clinical outcomes and may offer insight into targeted therapies.} 


Here, we concentrate on integrating gene expression and DNA methylation data. We downloaded the data from the TCGA website: https://cancergenome.nih.gov/   \citep[see also][]{international2010international} using the TCGA assembler 2 R package \citep{wei2017tcga, zhu2014tcga}. The gene expression profile is obtained using Illumina HiSeq Array, and the data were normalized counts of genes. The DNA methylation data, which contains percentage of methylation, is obtained using the Infinium Human Methylation 450 BeadChip. We acquired the level $3$ data for gene expression and DNA methlyation. We now briefly summarize the data pre-processing steps for both data as follows. We first filter out those genes with less than $5\%$ somatic mutation. Then we filter out the under-expressed genes with mean of  normalized expression less than $3.5$, and standard deviation  less than $1.7$. For the methylation data, we filter out the features for which the standard deviation of the percentages is less than $0.01$. Finally, clinical data is matched with samples that have both gene expression and DNA methylation data. After these steps, a common set of $88$ tumor samples is obtained, and the platforms consist of RNA gene expression data for $309$ genes and DNA methylation data for $288$ features.

We applied the model described in Section \ref{section: MEI} and the MCMC procedure detailed in Section \ref{section: PI} to obtain a global row clustering of patients with allowing platform-specific column clustering of probes, then to select informative cluster predictors by applying the variable selection procedure described in Section \ref{section: VS}. 


\paragraph{Identification of cancer subtypes} 

A point estimate of global row clusters were obtained via \cite{dahl2006model} to ease visualization and interpretation. Table \ref{table:cluster} and \ref{table:subtypes} respectively show matching matrices comparing the global row clusters given by our method with the cancer subtypes and transcriptional subtypes defined by TCGA.

Our method discovered different clustering structures than the subtypes defined by TCGA, but not independent (Fisher's exact test, p-value $\ll 0.0001$). {Clusters $1$, $2$ and $3$ primarily represent finer subpopulations of the LUAD samples which usually have a high rate of mutations. Cluster 1 corresponds to the transcriptional subtype proximal-inflammatory (PI), which was featured by co-mutation of tumour suppressor genes NF1 and TP53 as described in \cite{cancer2014comprehensive}. Cluster 2 and 3 are primarily subsets of proximal-proliferative (PP) that was enriched for mutation of KRAS. To examine chromatin states as suggested by \cite{cancer2014comprehensive}, a measure of methylation level at CpG island methylator phenotype (CIMP) sites is used. Majority proportions of cluster 1 and 2 ($75\%$ and $60\%$ respectively) correspond to DNA hypermethylation. Cluster 3 links to a more normal-like CIMP-low group (73\%). The prognosis for lung cancer is poor, our method reveals a well separation between samples that have different prognosis. Cluster 3 is formed by the samples with the worst 3-year survival rate of $0.08 \pm 0.07$, whereas cluster 4, mainly consists of the LUSC samples, has significantly better prognosis in 3 years, but with 6-year survival rate only $0.14 \pm 0.09$. The Kaplan-Meier survival curves for all global row clusters show significant difference (log-rank test p-value  $\ll 0.0001$). }

\begin{figure}[!pb]
\centering
	\includegraphics[width=84mm,height=5cm]{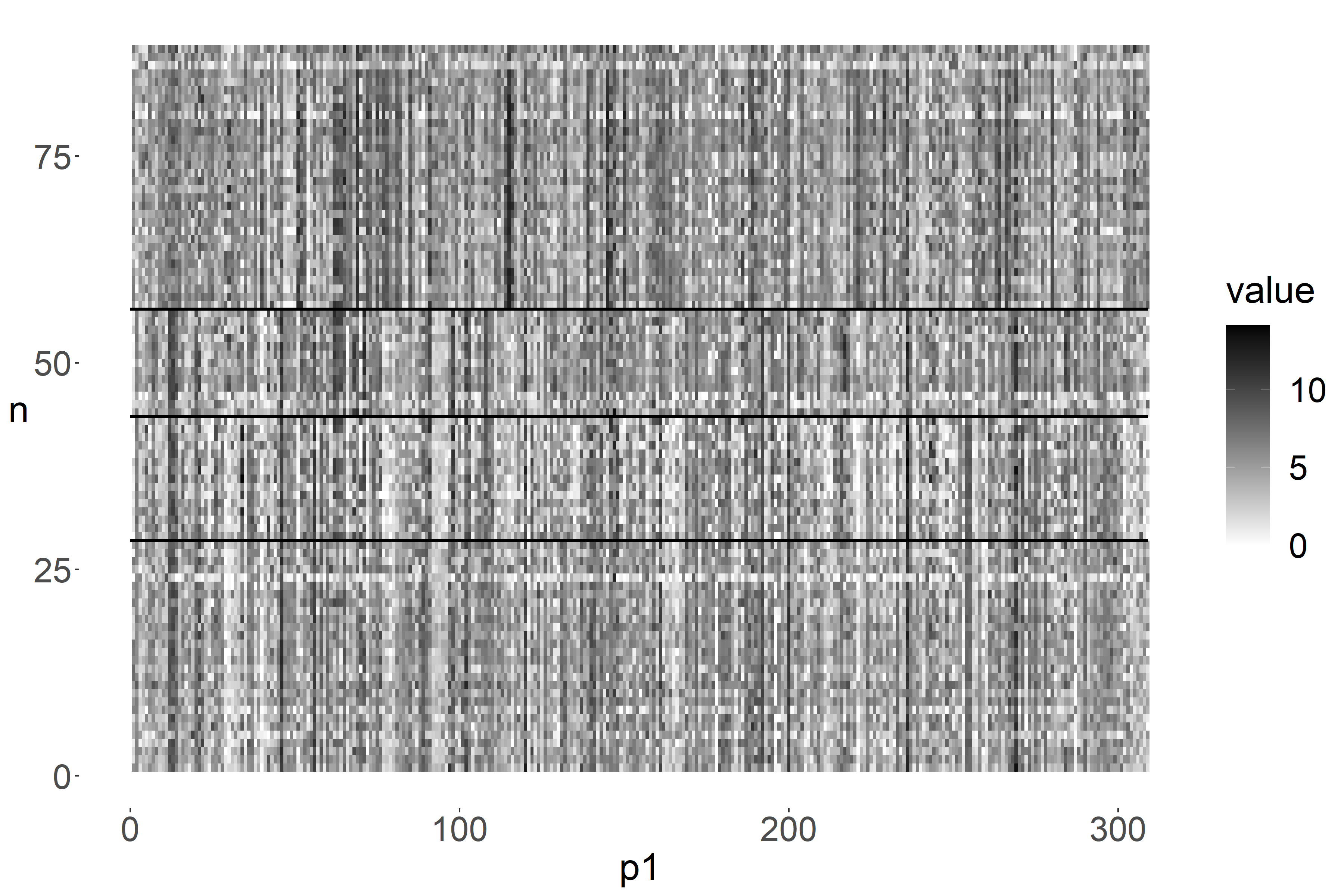}
	\includegraphics[width=84mm,height=5cm]{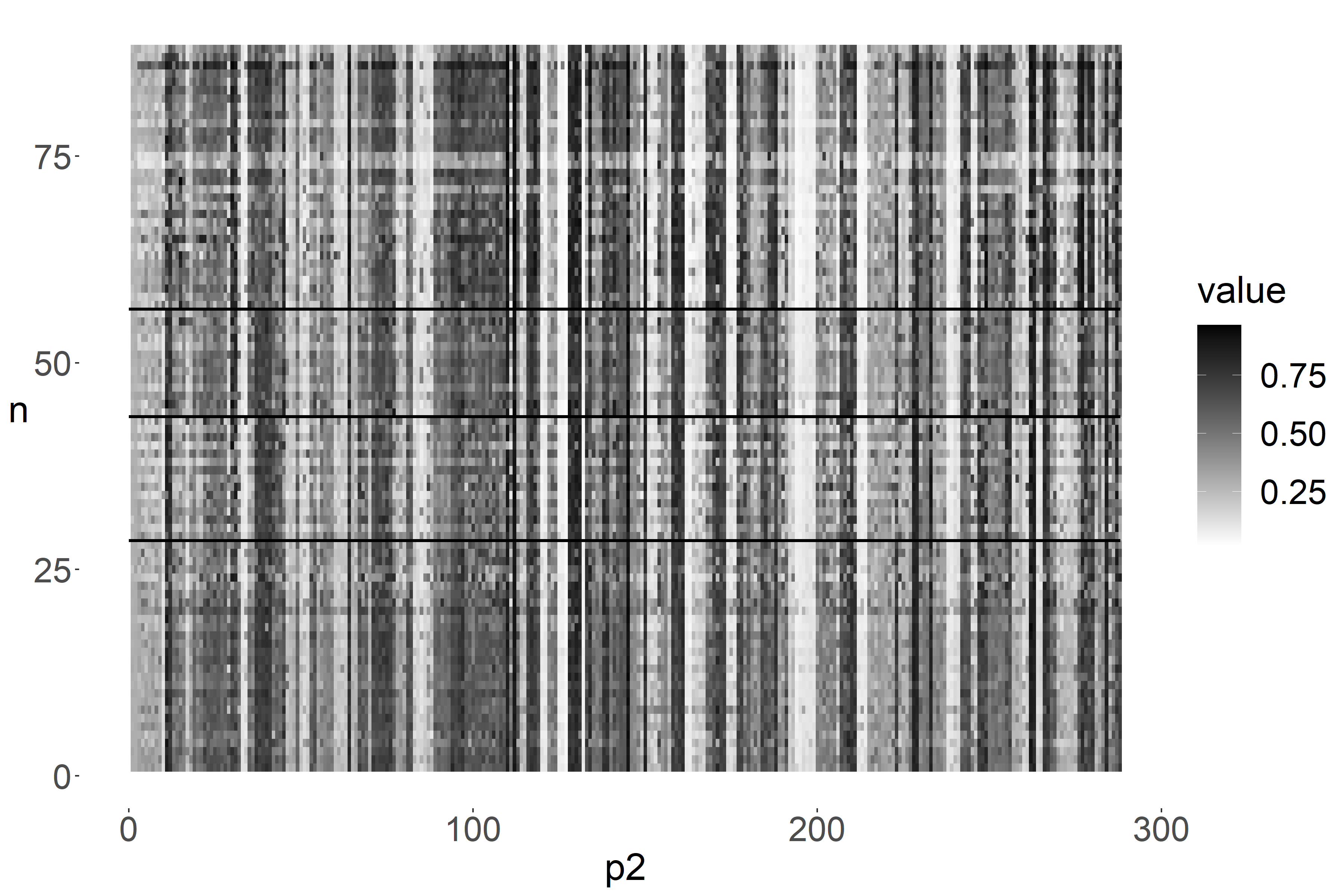}
	\caption[0.5\textwidth]{Heatmaps of gene expression (mRNA) and DNA methylation. Samples are arranged by both {\it row least-squares allocation} and {\it column least-squares allocation} under the proposed model.}
	\label{figure: mRNAmeth}
\end{figure}

Figure \ref{figure: mRNAmeth} shows the heatmap of mRNA and methylation with samples rearranged by their {\it row least-squares allocation} and probes rearranged by {\it column least-squares allocation}. The cluster are annotated in a descending order. We annotate the cluster at the bottom of each heatmap as cluster $1$, the cluster at the top of each heatmap as cluster $4$.

\paragraph{Selected features} Applying the gene selection technique described in Section \ref{section: VS}, and at Bayesian FDR = $0.2$, $64$ among the $239$ column clusters discovered in mRNA are selected as {cluster predictors}. Whereas, $14$ among $79$ column clusters discovered in methylation are selected as cluster predictors. {Among the $78$ cluster predictors detected from either mRNA or methylation platform, we list the top eight clusters that are most predictive of patients' survival outcomes in Table \ref{table:features}, with the posterior probabilities of being selected as predictors greater than 0.8. Since the cluster labels are arbitrary, we rank them in descending order of the posterior probabilities, with cluster $1$ as the most predictive. Each cluster contains one or multiple genes or CpG islands along with their posterior probabilities of being selected as a cluster representative, e.g.,   gene ORM1 has a posterior probability of $0.217$ of being the cluster representative of cluster $4$. } 

{These results confirm several findings in the scientific literature. \cite{relli2018distinct} discovered tumor prognostic drivers that have impact on LUSC or LUAD clinical outcome based on a next generation sequencing analysis of the NSCLC. Among the discovered determinants, SERPINB5 overexpression is related to dismal prognosis in LUAD. As the RNA level analysis in this paper extended to protein level, the protein coded by FGFBP1 gene is one of the key proteomic signatures in the prognosis profiles of either lung cancer subtype.  \cite{jiang2016breast} suggested several SNPs at the TOX3/LOC643714 locus that might promote lung cancer risk. The locus was previously identified through GWAS as one of the first regions associated with breast cancer. Recently, \cite{bauer2017epiregulin} discovered that lung tumor promotion was significantly reduced in the EREG knockout compared to wildtype controls in a murine model, suggesting that EREG could be used as a biomarker as well as a potential novel target in lung cancer, which largely overlaps with our findings. Another study aimed to understand the mechanisms behind epidermal growth factor receptor tyrosine kinase inhibitors (EGFR-TKIs) resistance further suggested that EREG, predominantly produced in macrophages in the tumor tissue, as a novel biomarker and moderator for EGFR-TKI therapeutics \citep{ma2021epiregulin}. }


\section{Discussion}\label{section: DIS}

In this article, we have proposed an innovative, flexible and scalable Bayesian nonparametric framework for analyzing multi-domain array and next generation sequencing-based omics datasets, and detecting various aspects of molecular drivers for multiple platforms. Different from most model based integrative analysis, which focus on either identifying biological associations of probes among different platforms or grouping subjects according to the similarities of their molecular entities, our model involves a nonparametric hierarchical structure, which estimates biological mechanisms by an effective bidirectional clustering to achieve dimension reduction at both probes and subjects level. Additionally, an efficient variable selection procedure is implemented to identify relevant  biological features that can potentially reveal underlying drivers of a disease. 

The major task of the model is to identify cancer subtypes. Through  simulation studies and comparisons, our model gives high clustering accuracy in global subject--to--row-cluster mappings and data platforms specific local column allocation of features. The application to TCGA data confirms the molecular profiling of LUSC and LUAD discussed in previous literature that may offer perception into targeted therapies. The proposed model provides an intuitive and insightful setting for integrative clustering of multiple domains to improve cancer subtype diagnosis and prognosis.



Even though the motivating data involves two platforms, the methodology proposed in this paper can be applied as is to an arbitrary number of platforms and disease subtypes. Generalization to count, categorical, and ordinal probes is possible. It is important to investigate the dependence structures and theoretical properties associated with the more general framework. This will be the focus of our group's future research. 

As a result of the complex nature of the MCMC inference, we performed these analyses in two stages, with bidirectional clustering followed by subtypes discovery, then predictor discovery. We are currently working on implementing the MCMC procedure in a parallel computing structure using graphical processing units (GPU) \citep{suchard2010understanding}. We plan to make the software available as an R package for general use in the near future. 

\section*{Acknowledgments}

This work was supported by the National Science Foundation under Award
DMS-1854003 to SG.

\bibliographystyle{natbib}
%
%
\bibliography{document.bib}

\newpage

	\begin{table}[!b]
	\centering
	\caption{Column 2 displays the mean proportion of correctly clustered probe pairs. Column 3 displays the estimated $R^2$, with the standard errors for the $50$ independent replications shown in parentheses. }
	\begin{tabular}{c c c c}
	\toprule
	 & & \textbf{Percent} $\hat{\varkappa}_{1t}$ & \textbf{Estimated} $R^2$ \\
	 \midrule
	                & $t=1$ &  0.971 (0.010) & 0.924 (0.038) \\
	                & $t=2$ &  0.975 (0.011) & 0.943 (0.025) \\
    \toprule
	\end{tabular}
	\label{table:column_percent}
	\end{table}

	\begin{table}[!tpb]
	\centering
	\small
	\caption[0.8\textwidth]{Global row clustering versus TCGA comprehensive subtype matching matrix\label{table:cluster}}
	\begin{tabular}{ccccc}
	\toprule
     & \multicolumn{4}{c}{Row cluster}  \\
    \cmidrule{2-5}	
       Cancer subtypes  & 1 & 2 & 3 & 4 \\
    	\midrule
    LUAD & 28 & 15 & 11 &  2 \\
    LUSC & 0 & 0 & 2 & 30 \\
    \midrule 
    3-year survival  & $0.36 \pm 0.09$ & $0.40 \pm 0.12$ & $0.08 \pm 0.07$ & $0.63 \pm 0.08$ \\ 
    \toprule
	\end{tabular}
	\end{table}
	
	    \begin{table}[!pb]
	    \centering
	\caption[0.8\textwidth]{First three global row clusters versus TCGA lung adenocarcinoma (LUAD) molecular subtypes matching matrix\label{table:subtypes}}
	\begin{tabular}{lccc}
	\toprule
     & \multicolumn{3}{c}{Row cluster}  \\
    \cmidrule{2-4}	
        Molecular subtypes of LUAD    & 1 & 2 & 3  \\
    	\midrule
    Proximal-inflammatory (PI) & 23 & 1 & 2  \\
    Proximal-proliferative (PP) & 0 & 9 & 8  \\
    Terminal respiratory unit (TRU) &  5 & 5 &  1  \\
    \midrule
    Mutation rate &&&\\
    NF1  & 32\% & 13\% & 9\%  \\
    TP53 & 75\% & 26\% & 36\% \\
    KRAS & 17\% & 33\% & 36\% \\
    \midrule
    Methylation level &&&\\
    CIMP-high  & 11\% & 40\% & 18\% \\
    CIMP-intermediate  & 64\% & 20\% & 9\% \\
	\toprule
	\end{tabular}
	\end{table}
	
		    \begin{table}[!pb]
	    \centering
	\caption[0.8\textwidth]{Top eight clusters of mRNA and methylation that are most predictive \label{table:features}}
	\begin{tabular}{l|cc|c}
	\toprule
    \cmidrule{2-4}	
        Cluster    & Platform & Gene or CpG island & Probability  \\
    	\midrule
    1 & mRNA  & FGFBP1  & 1  \\
       \midrule 
    2 & mRNA & SERPINB5 & 1 \\
         \midrule 
    3 & mRNA  & TOX3  & 1  \\
    	\midrule
     &   & ORM1 &  0.217  \\
     &   & VSIG2  &  0.213 \\
   4 & mRNA  & SFTA1P  & 0.185  \\
     &   & PLA2G10  &  0.156 \\
     &  & CFAP221 & 0.131  \\
     &  & COL4A3 & 0.099  \\
    \midrule
   5 & mRNA & EREG & 1  \\
   \midrule
    &   & SFRP1  & 0.150 \\
    &   & FOXA2  & 0.130 \\
    &   & TOX3  & 0.115 \\
    &   & PEG10  & 0.111 \\
  6 &  Methylation & CFAP221 & 0.110 \\
    &   & GFRA3  & 0.109 \\
    &   & KREMEN2  & 0.095 \\
    &   & NSG1  & 0.095 \\
    &   & CRYM & 0.087 \\
   \midrule
    &  & CNTN1 & 0.368 \\
  7 & mRNA & VSNL1 & 0.353 \\
    &  & GRHL3 & 0.279 \\
      \midrule
    &  & SAA2 & 0.362 \\
  8 & Methylation & CYP4B1 & 0.356 \\
    &  & MUC13 &  0.283 \\
	\toprule
	\end{tabular}
	\end{table}

\begin{figure}[!tpb]
 \centerline{\includegraphics[width=84mm,height=5cm]{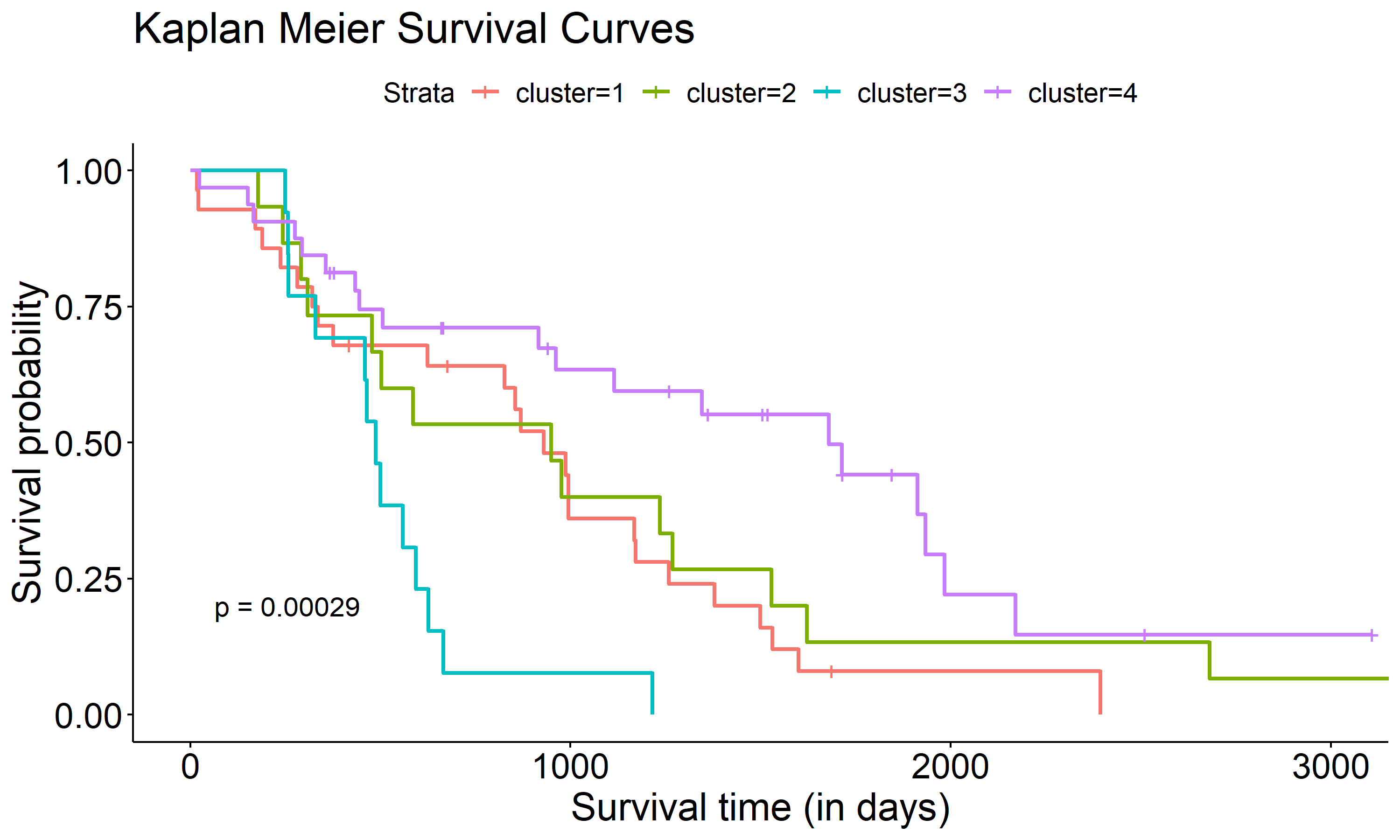}}
	\caption[0.7\textwidth]{Kaplan-Meier survival curves of discovered global row clusters for lung cancer specific survivals. Crosses indicate censored outcomes. }
	\label{figure: kmfit}
\end{figure}

\end{document}